\RequirePackage{ifpdf}
\ifpdf 
\documentclass[pdftex]{sigma}
\else
\documentclass{sigma}
\fi

\begin{document}


\renewcommand{\PaperNumber}{046}

\FirstPageHeading

\renewcommand{\thefootnote}{$\star$}

\ShortArticleName{Hamiltonian Systems Inspired by the Schr\"{o}dinger Equation}

\ArticleName{Hamiltonian Systems Inspired\\ by the Schr\"{o}dinger Equation\footnote{This paper is a
contribution to the Proceedings of the Seventh International
Conference ``Symmetry in Nonlinear Mathematical Physics'' (June
24--30, 2007, Kyiv, Ukraine). The full collection is available at
\href{http://www.emis.de/journals/SIGMA/symmetry2007.html}{http://www.emis.de/journals/SIGMA/symmetry2007.html}}}

\Author{Vasyl KOVALCHUK and Jan Jerzy S\L AWIANOWSKI}

\AuthorNameForHeading{V. Kovalchuk and J.J. S\l awianowski}

\Address{Institute of Fundamental Technological Research, Polish Academy of Sciences,\\
21, \'{S}wi\c{e}tokrzyska str., 00-049 Warsaw, Poland}
\Email{\href{mailto:vkoval@ippt.gov.pl}{vkoval@ippt.gov.pl}, \href{mailto:jslawian@ippt.gov.pl}{jslawian@ippt.gov.pl}}

\URLaddress{\url{http://www.ippt.gov.pl/~vkoval/}, \url{http://www.ippt.gov.pl/~jslawian/}}

\ArticleDates{Received October 30, 2007, in f\/inal form April 25,
2008; Published online May 27, 2008}

\Abstract{Described is $n$-level quantum system realized in the $n$-dimensional ``Hilbert'' space $H$ with the scalar product $G$ taken as a dynamical variable. The most general Lag\-rangian for the wave function and $G$ is considered. Equations of motion and conservation laws are obtained. Special cases for the free evolution of the wave function with f\/ixed $G$ and the pure dynamics
of $G$ are calculated. The usual, f\/irst- and second-order modif\/ied Schr\"{o}dinger equations are obtained.}

\Keywords{Schr\"{o}dinger equation; Hamiltonian systems on manifolds of scalar products; $n$-level quantum systems; scalar product as a dynamical variable; essential non-perturbative nonlinearity; conservation laws; GL$(n,\mathbb{C})$-invariance}

\Classification{81P05; 81R05; 81Q99; 37J05; 15A04; 15A63; 15A90; 20G20}

\section{Introduction}

Conf\/iguration spaces endowed with some algebraic structures are of interest in various areas
of mathematical physics. As a rule, Hamiltonian systems def\/ined on their cotangent bundles
have certain mathematically and physically interesting features, especially when their
Hamiltonians are somehow suited to the mentioned algebraic structures, e.g., are invariant
under their automorphism groups or subgroups. The best known example is the theory of
Hamiltonian systems on the cotangent bundles of Lie groups or their group spaces (or even more
general homogeneous spaces) where by the group space we mean the homogeneous space with
trivial isotropy groups, i.e., groups which ``forgot'' about having the distinguished neutral
element. The special attention in applications is paid to Hamiltonians invariant under left or
right translations or under both of them. The examples are the rigid bodies, incompressible
ideal f\/luids \cite{Arn78}, af\/f\/inely-rigid bodies (see for example \cite{all04,all05} and
references therein), etc.

Usually in physics one deals with linear groups, i.e., groups faithfully realizable by f\/inite
matrices. The only relatively known exceptions are $\overline{{\rm GL}(n,\mathbb{R})}$ and
$\overline{{\rm SL}(n,\mathbb{R})}$, i.e., the covering groups of ${\rm GL}(n,\mathbb{R})$ and
${\rm SL}(n,\mathbb{R})$ respectively. However, in spite of various attempts of F.~Hehl, Y.~Ne'eman and others (see for example~\cite{HKH,HN}), their physical applicability is as yet rather
doubtful and questionable. So, one of the best known examples of Hamiltonian systems on
algebraic structures are (usually invariant) ones on the cotangent bundles of matrix groups or
more gene\-ral\-ly some matrix manifolds. From the purely algebraic point of view, such
conf\/iguration spaces consist of second-order (and non-degenerate) tensors in some linear
spaces. Geometrically they represent linear transformations. Some questions appear here in a
natural way. Namely, it is a rule that all second-order tensors, i.e., not only mixed ones,
are of particular importance in physics. Twice covariant or contravariant tensors represent
various scalar products, e.g., metric tensors, electromagnetic f\/ields, gauge f\/ields, etc. In a
purely analytical sense all second-order tensors are matrices. Obviously, due to the
dif\/ference in the transformation rules, all they are geometrically completely dif\/ferent
objects. Nevertheless the natural question arises as to the existence of geometrically and
physically interesting Hamiltonian systems on the cotangent bund\-les of manifolds of
second-order tensors of other type than linear transformations. We mean here f\/irst of all the
manifolds of scalar products, both real-symmetric and complex-sesquilinear-hermitian. Also the
twice covariant and contravariant tensors without any special symmetries may be interesting.

One of our motivations has to do with certain ideas concerning nonlinear quantum mecha\-nics.
Various ways towards nonlinearity in quantum case were presented, e.g., in the review papers~\cite{Sv_rev,Sv}, from those motivated by paradoxes of the
quantum measurement, the interplay of unitary evolution and reduction,
etc., to certain ideas based on geometry like, for instance, the Doebner--Goldin nonlinearity \cite{DG,DGN,G}. However in
this article we are motivated by another idea. Namely, it is well known that the unitary
evolution of a quantum system, described by the Schr\"{o}dinger equation, may be interpreted
as a Hamiltonian system on Hilbert space. The most convenient way to visualize this is to
start from f\/inite-dimensional, i.e., ``$n$-level'', quantum systems~\mbox{($n<\infty$)}. The scalar
product is then f\/ixed once for all and is an absolute element of the system. The true ``degrees
of freedom'' are represented only by the vector of the underlying Hilbert space ``wave
functions''. And here some natural analogy appears with the situation in Special Relativity vs.\
General Relativity:
\begin{itemize}\itemsep=0pt
\item In specially-relativistic theories the metric tensor is f\/ixed once for all as an
absolute object, whereas all physical f\/ields are ``f\/lexible'' and satisfy dif\/ferential equations
as a rule derivable from the variational principle. The f\/ixed metric tensor is then used as a
``glue'' to contract tensor indices in order to build the scalar density of weight one dependent
algebraically on f\/ields and their f\/irst-order derivatives.

\item In generally-relativistic theories the metric tensor becomes f\/lexible as well, it is
included to degrees of freedom and satisf\/ies dif\/ferential equations together with the other
``physical'' f\/ields. Moreover it becomes itself the physical f\/ield, in this case the
gravitational one.
\end{itemize}

One can wonder whether one should not follow a similar pattern in quantum mechanics. Just to
make the scalar product ``f\/lexible'' and dynamically coupled to the $\psi$-object, i.e., to the
``wave function''. But, as mentioned, the scalar product is a twice covariant tensor. And so we
return to the idea of Hamiltonian systems on manifolds of scalar products or more general
twice covariant or twice contravariant tensors. And the point is that such manifolds carry
some natural Riemannian, pseudo-Riemannian or hermitian metric structures (almost canonical)
which are essentially non-Euclidean, i.e., describe some curved geometries on manifolds of
scalar products. Because of this the coef\/f\/icients at their derivatives in Lagrangians (as
quadratic forms of those velocities) are irreducibly non-constant. The resulting
Euler--Lagrange equations for them, and therefore also for the systems ``wave function $+$
f\/lexible scalar products'', are essentially nonlinear. This is the non-perturbative
nonlinearity, i.e., it cannot be interpreted as an artif\/icial extra correction to some basic
linear background. So, physically this is one of natural candidates for the ef\/fective and
geometrically interpretable nonlinearity in quantum mechanics, perhaps somehow explaining the
conf\/lict between unitary evolution and reduction, which exists essentially due to the
linearity of the standard quantum mechanics.

Beside the above-mentioned physical motivations, one should also stress that such Hamiltonian
models are interesting in themselves from the purely geometric point of view. They are somehow
similar to the (pseudo-)Riemannian metric structures on semisimple Lie groups, in particular
to the Killing tensor. Nevertheless their algebraic and geometric structure is dif\/ferent. As
to our knowledge, such Riemannian geometries have not been yet studied in mathematics. One has
the feeling that being so canonical as Killing metrics on groups they may have some
interesting geometric properties and are worth to be investigated.

\section{General problem}

Let us take a set of $n$ elements and some function $\psi$ def\/ined on it, i.e.,
\begin{gather*}
\mathcal{N}=\{1,\ldots,n\}\in \mathbb{N},\qquad \psi:\mathcal{N}\rightarrow\mathbb{C}.
\end{gather*}
Then we can def\/ine the ``wave function'' of the $n$-level quantum system as a following
$n$-vector
\begin{gather*}
\psi=\left[
\begin{array}{c}
  \psi^{1} \\
  \vdots \\
  \psi^{n}
\end{array}
\right],\qquad \psi^{a}=\psi(a)\in\mathbb{C}.
\end{gather*}
Let $H$ be a unitary space with the scalar product
\begin{gather*}
G:H\times H\rightarrow\mathbb{C},
\end{gather*}
which is a sesquilinear hermitian form. Then such an $H$ will be our $n$-dimen\-sional ``Hilbert''
space ($\mathbb{C}^{n}$).

So, let us consider the general Lagrangian
\begin{gather}
L=\alpha_{1}iG_{\bar{a}b}\big(\overline{\psi}{}^{\bar{a}}\dot{\psi}^{b}-
\dot{\overline{\psi}}{}^{\bar{a}}\psi^{b}\big)+
\alpha_{2}G_{\bar{a}b}\dot{\overline{\psi}}{}^{\bar{a}}
\dot{\psi}^{b}+
\big[\alpha_{4}G_{\bar{a}b}+\alpha_{5} H_{\bar{a}b}\big]
\overline{\psi}{}^{\bar{a}}\psi^{b}\nonumber\\
\phantom{L=}{} +\alpha_{3}\big[G^{b\bar{a}}+
\alpha_{9}\overline{\psi}{}^{\bar{a}}\psi^{b}\big]
\dot{G}_{\bar{a}b}+
\Omega[\psi,G]^{d\bar{c}b\bar{a}}
\dot{G}_{\bar{a}b}\dot{G}_{\bar{c}d}-\mathcal{V}\left(\psi,G\right),\label{eq8}
\end{gather}
where
\begin{gather*}
\Omega[\psi,G]^{d\bar{c}b\bar{a}}=
\alpha_{6}\big[G^{d\bar{a}}+\alpha_{9}\overline{\psi}{}^{\bar{a}}\psi^{d}
\big]\big[G^{b\bar{c}}+\alpha_{9}\overline{\psi}{}^{\bar{c}}\psi^{b}
\big]+
\alpha_{7}\big[G^{b\bar{a}}+\alpha_{9}\overline{\psi}{}^{\bar{a}}\psi^{b}
\big]\big[G^{d\bar{c}}+\alpha_{9}\overline{\psi}{}^{\bar{c}}\psi^{d}
\big]\nonumber\\
\phantom{\Omega[\psi,G]^{d\bar{c}b\bar{a}}=}{} +
\alpha_{8}\overline{\psi}{}^{\bar{a}}\psi^{b}
\overline{\psi}{}^{\bar{c}}\psi^{d},\qquad
\Omega[\psi,G]^{d\bar{c}b\bar{a}}=\Omega[\psi,G]^{b\bar{a}d\bar{c}},
\end{gather*}
and the potential $\mathcal{V}$ can be taken, for instance, in the following quartic form
\begin{gather*}
\mathcal{V}\left(\psi,G\right)=\varkappa\big(
G_{\bar{a}b}\overline{\psi}{}^{\bar{a}}\psi^{b}\big)^{2}.
\end{gather*}
The f\/irst and second terms in (\ref{eq8}) (those with
$\alpha_{1}$ and $\alpha_{2}$) describe the free
evolution of wave function $\psi$ while $G$ is f\/ixed.
The Lagrangian for trivial part of the linear dynamics
(those with~$\alpha_{4}$) can be also taken in the more general form
$f\left(G_{\bar{a}b}\overline{\psi}{}^{\bar{a}}\psi^{b}\right)$,
where $f:\mathbb{R}\rightarrow\mathbb{R}$.
The term with~$\alpha_{5}$ corresponds to the Schr\"{o}dinger
dynamics while $G$ is f\/ixed and then
\begin{gather*}
H^{a}{}_{b}=G^{a\bar{c}}H_{\bar{c}b}
\end{gather*}
is the usual Hamilton operator. If we properly choose the
constants $\alpha_{1}$ and $\alpha_{5}$, then we obtain precisely the Schr\"{o}dinger equation.
The dynamics of the scalar product $G$ is described by the terms linear and quadratic in the time derivative of $G$.
In the above formulae $\overline{\psi}{}^{\bar{a}}=\overline{\psi^{a}}$ denotes the usual
complex conjugation and $\alpha_{i}$, $i=\overline{1,9}$, and $\varkappa$ are some constants.

Then applying the variational procedure we obtain the equations of motion as follows
\begin{gather*}
\frac{\delta L}{\delta \overline{\psi}{}^{\bar{a}}}=
\alpha_{2}G_{\bar{a}b}\ddot{\psi}^{b}+\big(\alpha_{2}\dot{G}_{\bar{a}b}
-2\alpha_{1}iG_{\bar{a}b}\big)\dot{\psi}^{b}-
2\alpha_{8}\dot{G}_{\bar{a}b}\psi^{b}\dot{G}_{\bar{c}d}
\overline{\psi}{}^{\bar{c}}\psi^{d}
\nonumber\\
\phantom{{\delta \overline{\psi}{}^{\bar{a}}}=}{} -2\alpha_{9}\big(\alpha_{6}\dot{G}_{\bar{a}d}\dot{G}_{\bar{c}b}+
\alpha_{7}\dot{G}_{\bar{a}b}\dot{G}_{\bar{c}d}\big)
\psi^{b}\big(G^{d\bar{c}}+\alpha_{9}\overline{\psi}{}^{\bar{c}}\psi^{d}\big)
\nonumber\\
\phantom{{\delta \overline{\psi}{}^{\bar{a}}}=}{}
+\big[\big(2\varkappa G_{\bar{c}d}\overline{\psi}{}^{\bar{c}}\psi^{d}-
\alpha_{4}\big)G_{\bar{a}b}-\alpha_{5} H_{\bar{a}b}-
\big[\alpha_{3}\alpha_{9}+\alpha_{1}i\big]\dot{G}_{\bar{a}b}
\big]\psi^{b}=0
\end{gather*}
and
\begin{gather}
\frac{\delta L}{\delta G_{\bar{a}b}}=
2\Omega[\psi,G]^{b\bar{a}d\bar{c}}\ddot{G}_{\bar{c}d}+
2\dot{\Omega}[\psi,G]^{b\bar{a}d\bar{c}}\dot{G}_{\bar{c}d}+
\big(2\varkappa G_{\bar{c}d}\overline{\psi}{}^{\bar{c}}\psi^{d}-
\alpha_{4}\big)\overline{\psi}{}^{\bar{a}}\psi^{b}\nonumber\\
\phantom{{\delta G_{\bar{a}b}}=}{} +
2G^{d\bar{a}}\big[\alpha_{6}G^{b\bar{e}}\big(G^{f\bar{c}}+
\alpha_{9}\overline{\psi}{}^{\bar{c}}\psi^{f}\big)+
\alpha_{7}G^{b\bar{c}}\big(G^{f\bar{e}}+
\alpha_{9}\overline{\psi}{}^{\bar{e}}\psi^{f}\big)
\big]\dot{G}_{\bar{c}d}\dot{G}_{\bar{e}f}\nonumber\\
\phantom{{\delta G_{\bar{a}b}}=}{}-
\alpha_{2}\dot{\overline{\psi}}{}^{\bar{a}}\dot{\psi}^{b}+
\big[\alpha_{3}\alpha_{9}+\alpha_{1}i\big]
\dot{\overline{\psi}}{}^{\bar{a}}\psi^{b}+
\big[\alpha_{3}\alpha_{9}-\alpha_{1}i\big]
\overline{\psi}{}^{\bar{a}}\dot{\psi}^{b}=0,\label{eq20}
\end{gather}
where
\begin{gather*}
\dot{\Omega}[\psi,G]^{b\bar{a}d\bar{c}} =
\alpha_{8}\big(\dot{\overline{\psi}}{}^{\bar{a}}\psi^{b}
\overline{\psi}{}^{\bar{c}}\psi^{d}+\overline{\psi}{}^{\bar{a}}\dot{\psi}^{b}
\overline{\psi}{}^{\bar{c}}\psi^{d}+\overline{\psi}{}^{\bar{a}}\psi^{b}
\dot{\overline{\psi}}{}^{\bar{c}}\psi^{d}+\overline{\psi}{}^{\bar{a}}\psi^{b}
\overline{\psi}{}^{\bar{c}}\dot{\psi}^{d}\big)
\nonumber\\
\phantom{\dot{\Omega}[\psi,G]^{b\bar{a}d\bar{c}}=}{} +
\alpha_{6}\alpha_{9}\big(\big[\dot{\overline{\psi}}{}^{\bar{a}}\psi^{d}+
\overline{\psi}{}^{\bar{a}}\dot{\psi}^{d}\big]\big[G^{b\bar{c}}+
\alpha_{9}\overline{\psi}{}^{\bar{c}}\psi^{b}\big]+
\big[\dot{\overline{\psi}}{}^{\bar{c}}\psi^{b}+
\overline{\psi}{}^{\bar{c}}\dot{\psi}^{b}\big]\big[G^{d\bar{a}}+
\alpha_{9}\overline{\psi}{}^{\bar{a}}\psi^{d}\big]\big)
\nonumber\\
\phantom{\dot{\Omega}[\psi,G]^{b\bar{a}d\bar{c}}=}{} +
\alpha_{7}\alpha_{9}\big(\big[\dot{\overline{\psi}}{}^{\bar{a}}\psi^{b}+
\overline{\psi}{}^{\bar{a}}\dot{\psi}^{b}\big]\big[G^{d\bar{c}}+
\alpha_{9}\overline{\psi}{}^{\bar{c}}\psi^{d}\big]+
\big[\dot{\overline{\psi}}{}^{\bar{c}}\psi^{d}+
\overline{\psi}{}^{\bar{c}}\dot{\psi}^{d}\big]\big[G^{b\bar{a}}+
\alpha_{9}\overline{\psi}{}^{\bar{a}}\psi^{b}\big]\big)
\nonumber\\
\phantom{\dot{\Omega}[\psi,G]^{b\bar{a}d\bar{c}}=}{} -
\alpha_{6}\big[G^{d\bar{e}}G^{f\bar{a}}\big(G^{b\bar{c}}+
\alpha_{9}\overline{\psi}{}^{\bar{c}}\psi^{b}\big)+
G^{b\bar{e}}G^{f\bar{c}}\big(G^{d\bar{a}}+
\alpha_{9}\overline{\psi}{}^{\bar{a}}\psi^{d}\big)\big]\dot{G}_{\bar{e}f}
\nonumber\\
\phantom{\dot{\Omega}[\psi,G]^{b\bar{a}d\bar{c}}=}{} -
\alpha_{7}\big[G^{b\bar{e}}G^{f\bar{a}}\big(G^{d\bar{c}}+
\alpha_{9}\overline{\psi}{}^{\bar{c}}\psi^{d}\big)+
G^{d\bar{e}}G^{f\bar{c}}\big(G^{b\bar{a}}+
\alpha_{9}\overline{\psi}{}^{\bar{a}}\psi^{b}\big)\big]\dot{G}_{\bar{e}f}.
\end{gather*}

\section{Towards the canonical formalism}

The Legendre transformations leads us to the following canonical variables
\begin{gather}\label{eq01}
\pi_{b}=\frac{\partial L}{\partial \dot{\psi}^{b}}=
\alpha_{2}G_{\bar{a}b}\dot{\overline{\psi}}{}^{\bar{a}}+
\alpha_{1}iG_{\bar{a}b}\overline{\psi}{}^{\bar{a}},\qquad
\overline{\pi}_{\bar{a}}=
\frac{\partial L}{\partial \dot{\overline{\psi}}{}^{\bar{a}}}=
\alpha_{2}G_{\bar{a}b}\dot{\psi}{}^{b}-
\alpha_{1}iG_{\bar{a}b}\psi^{b},
\\
\label{eq02}
\pi^{\bar{a}b}=\frac{\partial L}{\partial \dot{G}_{\bar{a}b}}=
\alpha_{3}\big[G^{b\bar{a}}+
\alpha_{9}\overline{\psi}{}^{\bar{a}}\psi^{b}\big]+
2\Omega[\psi,G]^{b\bar{a}d\bar{c}}\dot{G}_{\bar{c}d}.
\end{gather}

The energy of our $n$-level Hamiltonian system is as follows
\begin{gather*}
E=\dot{\overline{\psi}}{}^{\bar{a}}\frac{\partial L}{\partial
\dot{\overline{\psi}}{}^{\bar{a}}}+\dot{\psi}^{b}\frac{\partial L}{\partial\dot{\psi}^{b}}+
\dot{G}_{\bar{a}b}\frac{\partial L}{\partial \dot{G}_{\bar{a}b}}-L\nonumber\\
\phantom{E}{} =
\alpha_{2}G_{\bar{a}b}\dot{\overline{\psi}}{}^{\bar{a}}
\dot{\psi}^{b}-
\left(\alpha_{4}
G_{\bar{a}b}+\alpha_{5}
H_{\bar{a}b}\right)\overline{\psi}{}^{\bar{a}}\psi^{b}
+\Omega[\psi,G]^{\bar{a}b\bar{c}d}
\dot{G}_{\bar{a}b}\dot{G}_{\bar{c}d}
+\varkappa\big(
G_{\bar{a}b}\overline{\psi}{}^{\bar{a}}\psi^{b}\big)^{2}.
\end{gather*}

Inverting the expressions (\ref{eq01}), (\ref{eq02}) we obtain that
\begin{gather*}
\dot{\overline{\psi}}{}^{\bar{a}}=
\frac{1}{\alpha_{2}}G^{b\bar{a}}\pi_{b}-
\frac{\alpha_{1}}{\alpha_{2}}i\overline{\psi}{}^{\bar{a}},\qquad
\dot{\psi}{}^{b}=\frac{1}{\alpha_{2}}G^{b\bar{a}}\overline{\pi}_{\bar{a}}+
\frac{\alpha_{1}}{\alpha_{2}}i\psi^{b},
\\
\dot{G}_{\bar{a}b}=
\frac{1}{2}\Omega[\psi,G]^{-1}_{\bar{a}b\bar{c}d}
\big(\pi^{\bar{c}d}-\alpha_{3}
\big[G^{d\bar{c}}+
\alpha_{9}\overline{\psi}{}^{\bar{c}}\psi^{d}\big]\big),
\end{gather*}
where
\begin{gather*}
\Omega[\psi,G]^{-1}_{\bar{a}b\bar{c}d}=
\Lambda[\psi,G]^{-1}_{\bar{a}b\bar{c}d}-
\frac{\alpha_{8}}{1+\alpha_{8}\theta_{2}[\psi,G]}
\Lambda[\psi,G]^{-1}_{\bar{a}b\bar{e}f}
\overline{\psi}{}^{\bar{e}}\psi^{f}
\Lambda[\psi,G]^{-1}_{\bar{c}d\bar{g}h}
\overline{\psi}{}^{\bar{g}}\psi^{h},\\
\Lambda[\psi,G]^{-1}_{\bar{a}b\bar{c}d}
=\frac{1}{\alpha_{6}}\lambda[\psi,G]^{-1}_{\bar{a}d}
\lambda[\psi,G]^{-1}_{\bar{c}b}-\frac{\alpha_{7}}{\alpha_{6}\left(
\alpha_{6}+n\alpha_{7}\right)}
\lambda[\psi,G]^{-1}_{\bar{a}b}\lambda[\psi,G]^{-1}_{\bar{c}d},\\
\lambda[\psi,G]^{-1}_{\bar{a}b}=
G_{\bar{a}b}-\frac{\alpha_{9}}{1+\alpha_{9}\theta_{1}[\psi,G]}
G_{\bar{a}d}G_{\bar{c}b}\overline{\psi}{}^{\bar{c}}\psi^{d},\\
\theta_{2}[\psi,G]=\Lambda[\psi,G]^{-1}_{\bar{a}b\bar{c}d}
\overline{\psi}{}^{\bar{a}}\psi^{b}
\overline{\psi}{}^{\bar{c}}\psi^{d}=\frac{
\alpha_{6}+\left(n-1\right)\alpha_{7}}{
\alpha_{6}\left(\alpha_{6}+n\alpha_{7}\right)}
\left(\frac{\theta_{1}[\psi,G]}{1+\alpha_{9}\theta_{1}[\psi,G]}\right)^{2},\\
\theta_{1}[\psi,G]= G_{\bar{a}b}\overline{\psi}{}^{\bar{a}}\psi^{b},
\end{gather*}
and then the Hamiltonian has the following form
\begin{gather*}
H=\frac{1}{\alpha_{2}}G^{b\bar{a}}\overline{\pi}_{\bar{a}}\pi_{b}
+\frac{\alpha_{1}}{\alpha_{2}}i\big(\psi^{b}\pi_{\psi b}-
\overline{\psi}{}^{\bar{a}}\overline{\pi}_{\bar{a}}\big)
-\left[\left(\alpha_{4}-\frac{\alpha^{2}_{1}}{\alpha_{2}}\right)G_{\bar{a}b}
+\alpha_{5}H_{\bar{a}b}\right]
\overline{\psi}{}^{\bar{a}}\psi^{b}\nonumber\\
\phantom{H=}{} +\frac{1}{4}\Omega[\psi,G]^{-1}_{\bar{a}b\bar{c}d}\pi^{\bar{a}b}\pi^{\bar{c}d}
-\frac{\alpha_{3}}{2}\Omega[\psi,G]^{-1}_{\bar{a}b\bar{c}d}\big[G^{b\bar{a}}+
\alpha_{9}\overline{\psi}{}^{\bar{a}}\psi^{b}\big]\pi^{\bar{c}d}\nonumber\\
\phantom{H=}{} +\frac{\alpha^{2}_{3}}{4}
\Omega[\psi,G]^{-1}_{\bar{a}b\bar{c}d}\big[G^{b\bar{a}}+
\alpha_{9}\overline{\psi}{}^{\bar{a}}\psi^{b}\big]
\big[G^{d\bar{c}}+
\alpha_{9}\overline{\psi}{}^{\bar{c}}\psi^{d}\big]
+\varkappa\big(
G_{\bar{a}b}\overline{\psi}{}^{\bar{a}}\psi^{b}\big)^{2}.
\end{gather*}

\section{Special cases}

\subsection[Pure dynamics for $G$]{Pure dynamics for $\boldsymbol{G}$}

First of all, if we consider the pure dynamics of scalar product $G$ while the wave function
$\psi$ is f\/ixed, then from (\ref{eq20}) we obtain the following equations of motion
\begin{gather}
\Omega[\psi,G]^{b\bar{a}d\bar{c}}\ddot{G}_{\bar{c}d}=
\left(\frac{\alpha_{4}}{2}-\varkappa \theta_{1}[\psi,G]\right)
\overline{\psi}{}^{\bar{a}}\psi^{b}+
\alpha_{7}\theta_{3}[\psi,G]
\big(G^{b\bar{a}}+
\alpha_{9}\overline{\psi}{}^{\bar{a}}\psi^{b}\big)
\nonumber\\
\phantom{\Omega[\psi,G]^{b\bar{a}d\bar{c}}\ddot{G}_{\bar{c}d}=}{} +
\alpha_{6}\dot{G}_{\bar{c}d}\dot{G}_{\bar{e}f}
\big(\gamma[\psi,G]^{b\bar{e}f\bar{c}d\bar{a}}
+\gamma[\psi,G]^{f\bar{a}d\bar{e}b\bar{c}}
-\gamma[\psi,G]^{b\bar{e}d\bar{a}f\bar{c}}\big),\label{eq25b}
\end{gather}
where
\begin{gather*}
\theta_{3}[\psi,G]=
G^{d\bar{e}}G^{f\bar{c}}\dot{G}_{\bar{c}d}\dot{G}_{\bar{e}f},\qquad
\gamma[\psi,G]^{f\bar{e}d\bar{c}b\bar{a}}=
G^{f\bar{e}}G^{d\bar{c}}\big(G^{b\bar{a}}+
\alpha_{9}\overline{\psi}{}^{\bar{a}}\psi^{b}\big).
\end{gather*}
If we additionally suppose that $\alpha_{4}=\alpha_{8}=\alpha_{9}=\varkappa=0$, then (\ref{eq25b}) simplif\/ies signif\/icantly
\begin{gather*}
\big(\alpha_{6}G^{b\bar{c}}G^{d\bar{a}}+
\alpha_{7}G^{b\bar{a}}G^{d\bar{c}}\big)
\big(\ddot{G}_{\bar{c}d}-
\dot{G}_{\bar{c}f}G^{f\bar{e}}\dot{G}_{\bar{e}d}\big)=0.
\end{gather*}
Hence, the pure dynamics of the scalar product is described by the following equations
\begin{gather}\label{eq25}
\ddot{G}_{\bar{a}b}-\dot{G}_{\bar{a}d}G^{d\bar{c}}\dot{G}_{\bar{c}b}=0.
\end{gather}
Let us now demand that $\dot{G}G^{-1}$ is equal to some constant value $E$, i.e., $\dot{G}=EG$, then
\begin{gather*}
\ddot{G}=E\dot{G}=E^{2}G
\end{gather*}
and
\begin{gather*}
\dot{G}G^{-1}\dot{G}=EGG^{-1}EG=E^{2}G,
\end{gather*}
therefore our equations of motion (\ref{eq25}) are fulf\/illed automatically and the solution is
as follows
\begin{gather*}
G(t)_{\bar{a}b}=\left(\exp(Et)\right)^{\bar{c}}{}_{\bar{a}}G_{0}{}_{\bar{c}b}.
\end{gather*}
Similarly if we demand that $G^{-1}\dot{G}$ is equal to some other constant $E^{\prime}$,
i.e., $\dot{G}=GE^{\prime}$,
\begin{gather*}
\ddot{G}=\dot{G}E^{\prime 2}=GE^{\prime 2},
\\
\dot{G}G^{-1}\dot{G}=GE^{\prime}G^{-1}GE^{\prime}=GE^{\prime 2},
\end{gather*}
then the equations of motion are also fulf\/illed and the solution is as follows
\begin{gather*}
G(t)_{\bar{a}b}=G_{0}{}_{\bar{a}d}\left(\exp(E^{\prime}t)\right)^{d}{}_{b}.
\end{gather*}
The connection between these two dif\/ferent constants $E$ and $E^{\prime}$ is written below
\begin{gather*}
\dot{G}(0)=\dot{G}_{0}=G_{0}E^{\prime}=EG_{0}.
\end{gather*}

\subsection[Usual and first-order modified Schr\"{o}dinger equations]{Usual and f\/irst-order modif\/ied Schr\"{o}dinger equations}

The second interesting special case is obtained when we suppose that the scalar product $G$ is
f\/ixed, i.e.,  the equations of motion are as follows
\begin{gather}\label{eq33}
\alpha_{2}\ddot{\psi}^{a}-2\alpha_{1}i\dot{\psi}^{a}+
\left(2\varkappa\theta_{1}\left[\psi,G\right]-
\alpha_{4}\right)\psi^{a}-\alpha_{5}H^{a}{}_{b}\psi^{b}=0.
\end{gather}
Then if we also take all constants of model to be equal to $0$ except of the following ones
\begin{gather*}
\alpha_{1}=\frac{\hbar}{2},\qquad \alpha_{5}=-1,
\end{gather*}
we end up with the well-known usual Schr\"{o}dinger equation
\begin{gather*}
i\hbar\dot{\psi}^{a}=H^{a}{}_{b}\psi^{b}.
\end{gather*}
Its f\/irst-order modif\/ied version is obtained when we suppose that $G$ is a dynamical variable
and~$\alpha_{2}$ is equal to $0$, i.e.,
\begin{gather}
i\hbar\dot{\psi}^{a} = H^{a}{}_{b}\psi^{b}-\left[\frac{i\hbar}{2}+\alpha_{3}\alpha_{9}\right]
G^{a\bar{c}}\dot{G}_{\bar{c}b}\psi^{b}+
\left(2\varkappa\theta_{1}\left[\psi,G\right]-
\alpha_{4}\right)\psi^{a}
\nonumber\\
\phantom{i\hbar\dot{\psi}^{a} =}{}-
2\alpha_{8}G^{a\bar{c}}\dot{G}_{\bar{c}b}\psi^{b}\dot{G}_{\bar{e}d}
\overline{\psi}{}^{\bar{e}}\psi^{d} -
2\alpha_{9}G^{a\bar{c}}\big(\alpha_{6}\dot{G}_{\bar{c}d}\dot{G}_{\bar{e}b}+
\alpha_{7}\dot{G}_{\bar{c}b}\dot{G}_{\bar{e}d}\big)
\psi^{b}\big(G^{d\bar{e}}+
\alpha_{9}\overline{\psi}{}^{\bar{e}}\psi^{d}\big),\!\!\!\label{eq37}\\
2\Omega[\psi,G]^{b\bar{a}d\bar{c}}\ddot{G}_{\bar{c}d}
=\left[\frac{i\hbar}{2}-\alpha_{3}\alpha_{9}\right]
\overline{\psi}{}^{\bar{a}}\dot{\psi}^{b}-
\left[\frac{i\hbar}{2}+\alpha_{3}\alpha_{9}\right]
\dot{\overline{\psi}}{}^{\bar{a}}\psi^{b}\nonumber\\
\phantom{2\Omega[\psi,G]^{b\bar{a}d\bar{c}}\ddot{G}_{\bar{c}d}=}{}-
2G^{d\bar{a}}\big[\alpha_{6}G^{b\bar{e}}\big(G^{f\bar{c}}+
\alpha_{9}\overline{\psi}{}^{\bar{c}}\psi^{f}\big)+
\alpha_{7}G^{b\bar{c}}\big(G^{f\bar{e}}+
\alpha_{9}\overline{\psi}{}^{\bar{e}}\psi^{f}\big)
\big]\dot{G}_{\bar{c}d}\dot{G}_{\bar{e}f}\nonumber\\
\phantom{2\Omega[\psi,G]^{b\bar{a}d\bar{c}}\ddot{G}_{\bar{c}d}=}{}-
\left(2\varkappa\theta_{1}\left[\psi,G\right]-
\alpha_{4}\right)\overline{\psi}{}^{\bar{a}}\psi^{b}-
2\dot{\Omega}[\psi,G]^{b\bar{a}d\bar{c}}\dot{G}_{\bar{c}d}.\nonumber 
\end{gather}
We can rewrite (\ref{eq37}) in the following form
\begin{gather*}
i\hbar\dot{\psi}^{a}=H_{\rm ef\/f}{}^{a}{}_{b}\psi^{b},
\end{gather*}
where the ef\/fective Hamilton operator is given as follows:
\begin{gather*}
H_{\rm ef\/f}{}^{a}{}_{b}
=
H^{a}{}_{b}-\left[\frac{i\hbar}{2}+\alpha_{3}\alpha_{9}\right]
G^{a\bar{c}}\dot{G}_{\bar{c}b}+
\left(2\varkappa\theta_{1}\left[\psi,G\right]-
\alpha_{4}\right)\delta^{a}{}_{b}-
2\alpha_{8}G^{a\bar{c}}\dot{G}_{\bar{c}b}\dot{G}_{\bar{e}d}
\overline{\psi}{}^{\bar{e}}\psi^{d}\nonumber\\
\phantom{H_{\rm ef\/f}{}^{a}{}_{b}=}{} -
2\alpha_{9}G^{a\bar{c}}\big(\alpha_{6}\dot{G}_{\bar{c}d}\dot{G}_{\bar{e}b}+
\alpha_{7}\dot{G}_{\bar{c}b}\dot{G}_{\bar{e}d}\big)\big(G^{d\bar{e}}+
\alpha_{9}\overline{\psi}{}^{\bar{e}}\psi^{d}\big).
\end{gather*}

\subsection[Second-order modified Schr\"{o}dinger equation]{Second-order modif\/ied Schr\"{o}dinger equation}

The idea of introducing the second time derivative of the wave function into the usual Schr\"{o}dinger equation as a correction term is not completely new and has been already discussed in the literature. The similar problems were studied a long time ago by A.~Barut and more recently have been re-investigated by V.V.~Dvoeglazov, S.~Kruglov, J.P.~Vigier and others (see, e.g.,~\cite{VVD,SIK} and references therein; the authors of this article are grateful to one of the referee for pointing them to above-mentioned references). Among others there is also an interesting article where the authors used the analogy between the Schr\"{o}dinger and Fourier equations~\cite{JMK}.

The quantum Fourier equation which describes the heat (mass) dif\/fusion on the atomic level has
the following form
\begin{gather*}
\frac{\partial T}{\partial t}=\frac{\hbar}{m}\nabla^{2}T.
\end{gather*}
If we make the substitutions $t\rightarrow it/2$ and $T\rightarrow \psi$, then we end up with
the free Schr\"{o}dinger equation
\begin{gather*}
i\hbar\frac{\partial \psi}{\partial t}=-\frac{\hbar^{2}}{2m}\nabla^{2}\psi.
\end{gather*}
The complete Schr\"{o}dinger equation with the potential term $V$ after the reverse
substitutions $t\rightarrow -2it$ and $\psi\rightarrow T$ gives us the parabolic quantum
Fokker--Planck equation, which describes the quantum heat transport for $\triangle t>\tau$, where $\tau=\hbar/m\alpha^{2}c^{2}\sim 10^{-17}$ sec and $c\tau\sim 1$~nm, i.e.,
\begin{gather*}
\frac{\partial T}{\partial t}=\frac{\hbar}{m}\nabla^{2}T-\frac{2V}{\hbar}T.
\end{gather*}
For ultrashort time processes when $\triangle t<\tau$ one obtains the generalized quantum
hyperbolic heat transport equation
\begin{gather*}
\tau\frac{\partial^{2} T}{\partial t^{2}}+\frac{\partial T}{\partial
t}=\frac{\hbar}{m}\nabla^{2}T-\frac{2V}{\hbar}T
\end{gather*}
(its structure and solutions for ultrashort thermal processes were investigated in
\cite{JMK_mono}) which leads us to the second-order modif\/ied Schr\"{o}dinger equation
\begin{gather}\label{eq56}
2\tau\hbar\frac{\partial^{2}\psi}{\partial t^{2}}+i\hbar\frac{\partial \psi}{\partial
t}=-\frac{\hbar^{2}}{2m}\nabla^{2}\psi+V\psi
\end{gather}
in which the additional term describes the interaction of electrons with surrounding
space-time f\/illed with virtual positron-electron pairs. It is easy to see that (\ref{eq56}) is
analogous to (\ref{eq33}) if we suppose that
\begin{gather*}
\alpha_{1}=\frac{\hbar}{2},\qquad \alpha_{2}=-2\tau\hbar,\qquad \alpha_{4}=0,\qquad
\alpha_{5}=-1,\qquad \varkappa=0.
\end{gather*}

\section[Conservation laws and GL$(n,\mathbb{C})$-invariance]{Conservation laws and GL$\boldsymbol{(n,\mathbb{C})}$-invariance}

So, if we investigate the invariance of our general Lagrangian (\ref{eq8})
under the group GL$(n,\mathbb{C})$ and consider some one-parameter group of transformations
\begin{gather*}
\left\{\exp\left(A\tau\right):\tau\in\mathbb{R}\right\},\qquad A\in {\rm L}(n,\mathbb{C}),
\end{gather*}
then the inf\/initesimal transformations rules for $\psi$ and $G$ are as follows
\begin{gather*}
\psi^{a} \mapsto L^{a}{}_{b}\psi^{b},\qquad 
G^{a\bar{c}} \mapsto L^{a}{}_{b}\overline{L}{}^{\bar{c}}{}_{\bar{e}}G^{b\bar{e}},
\qquad
G_{\bar{a}b} \mapsto G_{\bar{c}d}\overline{L^{-1}}{}^{\bar{c}}{}_{\bar{a}}L^{-1d}{}_{b},
\end{gather*}
where
\begin{gather*}
L^{a}{}_{b}=\delta^{a}{}_{b}+\epsilon A^{a}{}_{b},\qquad L^{-1a}{}_{b}\approx
\delta^{a}{}_{b}-\epsilon A^{a}{}_{b},\qquad \epsilon\approx 0.
\end{gather*}
So leaving only the f\/irst-order terms with respect to $\epsilon$ we obtain that the variations
of $\psi$ and $G$ are as follows
\begin{gather*}
\delta\psi^{a}=\epsilon A^{a}{}_{b}\psi^{b},\qquad  \delta\overline{\psi}{}^{\bar{a}}=\epsilon
\overline{A}{}^{\bar{a}}{}_{\bar{c}}\overline{\psi}{}^{\bar{c}},\\ 
\delta G^{a\bar{c}}=\epsilon\big(A^{a}{}_{b}G^{b\bar{c}}+
\overline{A}{}^{\bar{c}}{}_{\bar{e}}G^{a\bar{e}}\big),\qquad \delta
G_{\bar{a}b}=-\epsilon\big(G_{\bar{c}b}\overline{A}{}^{\bar{c}}{}_{\bar{a}}+
G_{\bar{a}d}A^{d}{}_{b}\big),
\end{gather*}
then
\begin{gather}\label{eq44}
\frac{1}{\epsilon}\left(\frac{\partial L}{\partial \dot{\overline{\psi}}{}^{\bar{a}}}\delta
\overline{\psi}{}^{\bar{a}}+\frac{\partial L}{\partial
\dot{\psi}^{b}}\delta\psi^{b}\right)=G_{\bar{a}b}
\big(\alpha_{2}\dot{\overline{\psi}}{}^{\bar{a}}+
\alpha_{1}i\overline{\psi}{}^{\bar{a}}\big)A^{b}{}_{d}\psi^{d}
+G_{\bar{a}b}\big(\alpha_{2}\dot{\psi}^{b}-\alpha_{1}i\psi^{b}\big)
\overline{A}{}^{\bar{a}}{}_{\bar{c}}\overline{\psi}{}^{\bar{c}}
\end{gather}
and
\begin{gather}
\frac{1}{\epsilon}\frac{\partial L}{\partial\dot{G}_{\bar{a}b}}\delta
G_{\bar{a}b}=-\big[\alpha_{3}\big(\delta^{b}{}_{f}+
\alpha_{9}G_{\bar{a}f}\overline{\psi}{}^{\bar{a}}\psi^{b}\big)
+2\Omega[\psi,G]^{b\bar{a}d\bar{c}}
G_{\bar{a}f}\dot{G}_{\bar{c}d}\big]A^{f}{}_{b}\nonumber\\
\hphantom{\frac{1}{\epsilon}\frac{\partial L}{\partial\dot{G}_{\bar{a}b}}\delta
G_{\bar{a}b}=}{}
 -\big[\alpha_{3}\big(\delta^{\bar{a}}{}_{\bar{e}}+
\alpha_{9}G_{\bar{e}b}\overline{\psi}{}^{\bar{a}}\psi^{b}\big)
+2\Omega [\psi,G ]^{b\bar{a}d\bar{c}}
G_{\bar{e}b}\dot{G}_{\bar{c}d}\big]\overline{A}{}^{\bar{e}}{}_{\bar{a}}.\label{eq45}
\end{gather}
If we consider some f\/ixed scalar product $G_{0}$ and take the $G_{0}$-hermitian $A$'s, then
\begin{gather*}
A^{a}{}_{b}=G_{0}{}^{a\bar{c}}\widetilde{A}_{\bar{c}b},\qquad
\overline{A}^{\bar{a}}{}_{\bar{c}}=\widetilde{A}_{\bar{c}b}G_{0}^{b\bar{a}},\qquad \widetilde{A}{}^{\dag}=\widetilde{A},
\end{gather*}
and therefore the expressions (\ref{eq44}) and (\ref{eq45}) are written together in the matrix form as follows
\begin{gather*}
\mathcal{J}\left(A\right)={\rm Tr}\big(V\widetilde{A}\big),
\end{gather*}
where the hermitian tensor $V$ describing the system of conserved physical quantities is given as follows
\begin{gather*}
V=\alpha_{2}\big(\psi\dot{\psi}^{\dag}GG^{-1}_{0}+G^{-1}_{0}G\dot{\psi}\psi^{\dag}\big)
+\big(\alpha_{1}i-\alpha_{3}\alpha_{9}\big)\psi\psi^{\dag}GG^{-1}_{0}\nonumber\\
\phantom{V=}{} -
\big(\alpha_{1}i+\alpha_{3}\alpha_{9}\big)G^{-1}_{0}G\psi\psi^{\dag}
-2\alpha_{3}G^{-1}_{0}
-2\big(G^{-1}_{0}G\omega [\psi,G ]+\omega [\psi,G ] GG^{-1}_{0}\big),
\end{gather*}
where
\begin{gather*}
\omega\left[\psi,G\right]^{b\bar{a}}=
\Omega\left[\psi,G\right]^{b\bar{a}d\bar{c}}\dot{G}_{\bar{c}d}.
\end{gather*}
Similarly for the $G_{0}$-antihermitian $A$'s, i.e., when $\widetilde{A}^{\dag}=-\widetilde{A}$, we
obtain another hermitian tensor~$W$ as a conserved value
\begin{gather*}
\mathcal{J}\left(A\right)={\rm Tr}\big(iW\widetilde{A}\big),
\end{gather*}
where
\begin{gather*}
iW = \alpha_{2}\big(\psi\dot{\psi}^{\dag}GG^{-1}_{0}-
G^{-1}_{0}G\dot{\psi}\psi^{\dag}\big)
+ (\alpha_{1}i-\alpha_{3}\alpha_{9} )\psi\psi^{\dag}GG^{-1}_{0}
\nonumber\\
\phantom{iW =}{} +
 (\alpha_{1}i+\alpha_{3}\alpha_{9} )G^{-1}_{0}G\psi\psi^{\dag}
+2\big(G^{-1}_{0}G\omega [\psi,G ]-\omega [\psi,G ] GG^{-1}_{0}\big).
\end{gather*}

\section{Final remarks}

This is a very preliminary, simplif\/ied f\/inite-level model. It is still not clear whether it is consistent with the usual statistical interpretation of quantum mechanics. This model is thought on as a~step towards discussing the wave equations obtained by combining the f\/irst and second time derivatives. There are some indications that such a combination might be reasonable. Within a rather dif\/ferent context (motivated by the idea of conformal invariance) we studied such a~problem in \cite{GF,KGD} where the wave equations with the superposition of Dirac and d'Alembert operators were considered.

\subsection*{Acknowledgements}

This paper contains results obtained within the framework of the research project 501 018 32/1992 f\/inanced from the Scientif\/ic Research Support Fund
in 2007--2010. The authors are greatly indebted to the Ministry of Science and Higher Education for this f\/inancial support.

The authors are also very grateful to the referees for their valuable remarks and comments concerning this article.

\pdfbookmark[1]{References}{ref}
\LastPageEnding
\end{document}